\documentclass[preprint,showpacs,preprintnumbers,amsmath,amssymb]{revtex4}

\usepackage{epsfig}
\usepackage{dcolumn}
\usepackage{bm}

\begin{document}

\title{Stress Relaxation in Aging Soft Colloidal Glasses}

\affiliation{Raman Research Institute, Bangalore 560080, INDIA}
\author{Ranjini Bandyopadhyay}
\email{ranjini@rri.res.in}
\author{P. Harsha Mohan}
\affiliation{Raman Research Institute, Bangalore 560080, INDIA}
\author{Yogesh M. Joshi}
\email{joshi@iitk.ac.in}
\affiliation{Department of Chemical Engineering, Indian Institute of Technology
Kanpur, Kanpur 208016, INDIA}
\vspace{0.5cm}

\date{\today}

\pacs{82.70.Dd, 61.20.Lc, 83.50.Ax, 83.80.Hj}
\maketitle

{\bf We investigate the stress relaxation behavior on the application of step strains to
aging aqueous suspensions of the synthetic clay Laponite. The stress exhibits a two-step decay, from which the slow relaxation modes are extracted as functions of the sample ages and applied step strain deformations. Interestingly, the slow time scales that we estimate show a dramatic enhancement with increasing strain amplitudes.  We argue that the system ends up exploring the deeper sections of its energy
landscape following the application of the step strain. }



\section{Introduction}
\vspace{1cm}
Concentrated suspensions, emulsions and gels exhibit rich
physical behavior and are widely used in the industry. Generically, these materials
are characterized by slow dynamics and often fail to achieve equilibrium within time
scales that can be achieved in the laboratory \cite{sol_prl}. They are
therefore classified as soft glassy materials and serve as excellent model systems
in the understanding of the non-equilibrium glass transition in hard condensed
matter systems \cite{cip_jpcm,ran_ssc}. As a direct consequence of the particle
crowding that exists in these materials, the constituent entities are
arrested in a cage formed by their neighbors. Such a jammed state limits their
diffusive movement beyond the cage dimensions, leading to an incomplete relaxation
process and a restricted access to the phase space \cite{cip_jpcm}. The
relaxation phenomena of such systems can be studied by analyzing the temporal behaviors of the
response functions, which typically exhibit non-exponential decays
\cite{deb_nature}. In this paper, we study the relaxation behavior of the shear stress upon the application of a step strain to aqueous
suspensions of Laponite, a model glass former. We observe that the stress decays in two
steps, with the slow time scale showing a super-linear dependence on age.  A two-step stress decay is generally observed for entangled polymeric melts \cite{doi_pol}. 
However, to the best of our knowledge, this is the
first reported rheological signature of a two-step relaxation of the shear stress in soft glassy materials. Contrary to
expectation, the slow relaxation mode shows an enhancement in sluggishness with the increase
in the magnitude of the step strain. We understand this observation in terms of strain-induced alterations of the system microstructure.

Laponite clay is composed of monodisperse, disc shaped particles of diameter 25 {\it
nm} and thickness 1 {\it nm}. It is generally observed that for concentrations above 1 volume \%, an aqueous suspension of Laponite undergoes ergodicity breaking soon after mixing Laponite powder in water \cite{bon_epl,ran_prl}. During the last decade, various
optical tools have been employed to investigate such ergodicity breaking and the subsequent aging dynamics in this system \cite{ran_ssc}.  The intensity autocorrelation function measured in dynamic light scattering (DLS) experiments is seen to exhibit a two-step decay. Such data is then analyzed to decouple the fast ($\beta$) and the
slow ($\alpha$) relaxation time scales of the suspension \cite{ran_prl,sch_pre,tan_pre2,abo_pre,bel_pre,ruz_prl}.

In a non-ergodic state, each Laponite disc can be considered to be trapped in the potential energy well created by its neighbors \cite{sol_prl}. In such an
arrested state, Laponite particles can undergo structural rearrangements that are driven by the local strain fields, thereby attaining a progressively lower potential
energy state. This results in an evolution of the properties of the system with time, a phenomenon commonly referred to as aging. The aging dynamics exhibited by soft glasses is strongly influenced by the deformation field \cite{dileo_pre,abou_jrheol,cloit_prl,der_pre,red_jap}. Sollich {\it et al.} \cite{sol_prl} proposed that the application of strain enhances the potential energy of the trapped particle, and if the barrier
height (or yield energy) of the energy well is exceeded, the particle escapes the well (or cage) and an yielding event occurs. Subsequent to the
yielding event, the particle gets trapped in a new cage and the aging dynamics is re-initialized. This process, wherein all the loading, shear and aging history of the sample is erased, is commonly referred to as rejuvenation \cite{ran_ssc,cloit_prl}.

\section{Sample Preparation and Viscometry}
Laponite RD, a synthetic hectorite clay used in this study, was procured from
Southern Clay Products, Inc. Powdered Laponite was dried overnight at 80$^{\circ}$C
before mixing it with deionized and distilled water. A basic pH ($\sim$10) was
maintained by the dropwise addition of a 1 mM solution of NaOH. The suspension was
stirred vigorously for 90 mins. Subsequently, the sample was filtered through a
Millipore Millex-HV 0.45 $\mu$m filter unit to ensure breakup of all the large
clusters in the suspension. All the rheological experiments reported in this work
were performed in an Anton Paar MCR 501 rheometer. The shear cell used was a double
gap geometry (outer diameter = 13.796 mm, inner diameter = 11.915 mm, cell height = 42 mm, sample volume required = 3.8 ml.)
 equipped with a water circulation unit for temperature control. All
experiments reported in this paper were performed at 20$^{\circ}$C and for a
Laponite concentration of 3.5 wt.\%. For each experimental run, the sample was
filled in the geometry and left to age for 90 minutes. In order to remove
experimental artifacts arising out of the sample preparation and loading process, the
suspensions were subjected to an oscillatory stress of amplitude 40 Pa at an angular
frequency of 0.1 rad/s. As expected, the suspensions yielded under these high stresses and
eventually showed a plateau of low viscosity that did not change with time. The
shear melting experiment was now stopped and the age of the sample $t_W$ was
measured from this time. The aging of the suspension was studied by recording the
evolution of the mechanical moduli of the suspension in oscillatory shear
experiments performed in the linear rheological regime.
\begin{figure}
\begin{center}
\includegraphics[width=3.5in]{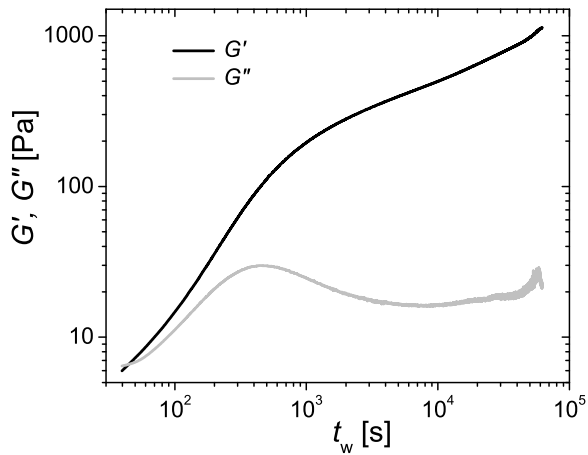}
\caption{Example data showing the evolution of the elastic and viscous moduli $G^{\prime}$ and $G^{\prime\prime}$ (black and grey solid lines respectively) with waiting time $t_{W}$.  Shear strains 
of amplitude 0.5\% at a frequency of 0.1 Hz were applied.\\}
\label{Fig 1}
\end{center}
\end{figure}

\section{Results and Discussion}
A typical aging behavior, showing the evolution of the elastic modulus G$^{\prime}$ (black line) and the viscous modulus G$^{\prime\prime}$ (grey line) as the
sample ages, is plotted in fig. 1. We continued such aging experiments until a
predetermined value of t$_W$ was reached. Subsequently, a step strain was applied to
the sample and the relaxation of stress was recorded. To ensure that wall slip did
not affect the acquired data, we repeated some of the experiments in a quartz
couette having a single gap. Results obtained using both the geometries coincided within experimental
errors.
\begin{figure}
\begin{center}
\includegraphics[width=3.5in]{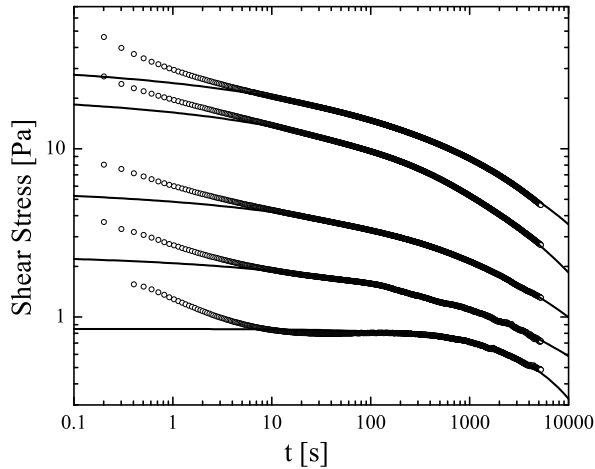}
\caption{Example data showing the decay of the shear stresses (circles) at different strains (from top to bottom: 100, 300, 2000, 5000 and 12000\%) and the corresponding fits of the slow mode (solid lines) to the KWW model. The step strains were applied to Laponite suspensions, each aged to 7200 s.   \\}
\label{Fig 2}
\end{center}
\end{figure}
Fig. 2 shows the stress relaxation behaviors exhibited by laponite
suspensions, of t$_W$ = 7200 seconds, on the application of step strains between 100\% (top-most curve) and 12000\% (lowest curve).  The stress relaxes in two steps showing two well-separated relaxation regimes. The stress decays rapidly within the first few seconds. Since our rate of data acquisition is limited to 10 data points/s, the statistics of the decay of the fast relaxation is not adequate, nor is its dynamic range broad enough, for us to model the fast relaxation process satisfactorily. The slower mode of the  stress relaxation function, however, can be modeled by the
Kohlrausch-Williams-Watts (KWW) function \cite{kww_paper,scher_rel}. The KWW function,
which describes the time-dependent stress relaxation $\sigma(t)$ of the slower mode in glassy materials
\cite{deb_nature}, is expressed as a stretched exponential term written as:
\begin{equation}
\sigma (t) = \phi\exp[-(t/t_{1})^B] 
\end{equation}
where $\phi$ is the baseline of the fast relaxation mode, $t_1$ is the time scale of the slow relaxation process and $B$ is a
stretching exponent between 0 and 1.  We fit the slower part of the stress relaxation curve to eq. 1 (solid lines in fig. 2) and
estimate the slow relaxation mode of the system for several suspension
ages and step strain magnitudes. Interestingly, intensity autocorrelation functions
measured in DLS experiments on aging Laponite suspensions also exhibit two stage
decays, with the decay of the slower process described well by eq. 1 \cite{bel_pre,ruz_prl}. This demonstrates the suitability
of the KWW equation in modeling the slow dynamics in glassy systems. The fast relaxation time scale
obtained in these experiments is identified with the rattling motion of the trapped entity within the cage formed
by its neighbors  \cite{ran_prl,sch_pre,tan_pre2}. The slow mode extracted from the KWW fits, on the other
hand, represents the relaxation of a particle as it hops between cages
\cite{bel_pre,sch_pre}.
\begin{figure}
\begin{center}
\includegraphics[width=3.5in]{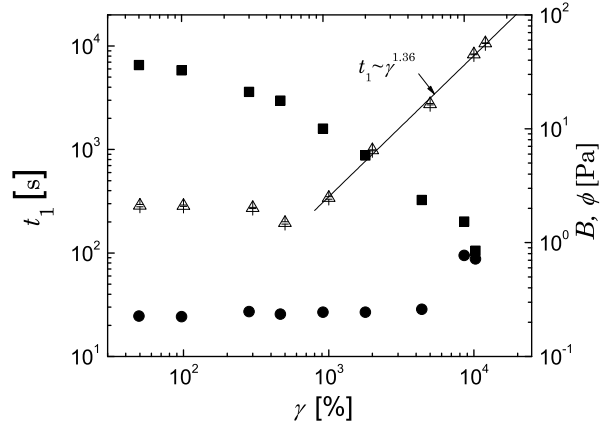}
\caption{ The slow relaxation mode $t_1$ (open triangles), the baseline of the fast relaxation process $\phi$ (filled squares) and the stretching exponent B (filled circles) are plotted against the
magnitude of the step strain $\gamma$ applied to Laponite suspensions of age 7200
s.\\}
\label{Fig 3}
\end{center}
\end{figure}


Fig. 3 shows the plot of the slow relaxation time $t_1$ versus applied strain deformation $\gamma$. Interestingly, the values of $t_1$ estimated from our experiments do not change with strain amplitudes up to 1000\%. Beyond this
value, it shows a power law increase with $\gamma$ that can be described by the
relation $t_1 \sim \gamma^{1.36 \pm 0.05}$. As larger strain amplitudes should 
activate the particle to overcome the energy barrier of the trap \cite{sol_prl}, thereby
speeding up its diffusion between cages and causing rejuvenation, our observation that the cage diffusion process actually slows down on the
application of a step strain is indeed quite counter-intuitive \cite{liu_nature}. Instead of facilitating hopping events between cages by
increasing the potential energy of the particles, we find here that the applied step strain actually results in the particle getting trapped in a deeper potential energy well. The decay time $t_{1}$ obtained from each long-time, stretched exponential relaxation corresponds to an average value in a situation presumably characterized by a wide distribution of relaxation times. While it is impossible for us to determine the complete relaxation time spectrum, our data for the average relaxation time $t_{1}$ exhibits the dramatic slow-down in the dynamics due the application of the step strains. Fig. 3 also describes the variation of the stretching parameter B (solid circles). It can be seen that for the lower step strains, B lies between 0.2 and 0.3. Clearly, owing to the large distribution of time scales that characterize these relaxation processes, a complete decay is not observed. However, at the highest
strain values achieved in our experiments, B approaches unity, indicating the possible onset of ordering into a nematic phase \cite{lele_jor}.
At even higher strains beyond 12000\%, the shape of the stress relaxation changes dramatically and cannot be modeled by the KWW function. While the relaxation moduli decrease sharply at these high strains, we cannot rule out wall slip completely under these conditions.  

Fig. 3 also shows that the baseline value ($\phi$) of the fast relaxation regime decreases as a function of the step strain amplitude. This behavior is also apparent from fig. 2, where the stress relaxation curve corresponding to a higher step strain lies below the stress relaxation curve corresponding to a lower step strain. A decrease in the modulus of the suspension could arise due to its partial rejuvenation upon the application of a step strain. However, this apparently contradicts our observation of the slowing down of the slow mode as the magnitude of the step strain is increased. We will discuss the possible reasons for this observation later in this section. From fig. 2, it can be seen that the fast relaxation process decays as a power law, indicating a distribution of relaxation times. 
The time required for the fast process to decay, which can be roughly estimated to be 1-2 s, is approximately independent of $\gamma$. Previous DLS studies \cite{abo_pre,sch_pre} that identified the fast relaxation time with the $\beta$ relaxation process have measured values that are significantly smaller than those estimated from our stress relaxation experiments. It is well-documented in the literature that different relaxing properties lead to estimations of relaxation times that could differ by more than two orders of magnitude \cite{scher_rel}.
In addition, stress is a macroscopic property, while intensity autocorrelation measurements performed in DLS experiments explore much smaller length scales. The two techniques are therefore sensitive to the system response at very different time scales due to the cooperativity involved in the former case. We believe that these factors are partially responsible for the observed discrepancy. 

We also performed oscillatory strain sweep experiments at a constant angular frequency. In contrast to the data reported above, we observed strain thinning with a power-law decrease of the elastic and viscous moduli above a certain critical strain value (fig. 4). This is an important observation, as it emphasizes that the slowing down of the cage diffusion process with increasing deformation is more specific to a scenario involving the application of strains over very short periods of time.
\begin{figure}
\begin{center}
\includegraphics[width=3.5in]{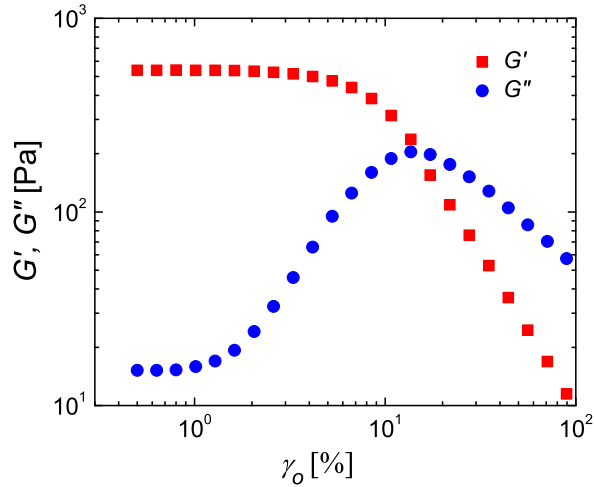}
\caption{ The evolution of the elastic (squares) and viscous moduli (circles) with an oscillatory strain of amplitude $\gamma_{\circ}$, applied at a fixed angular frequency of 1 rad/s. The age of the sample was $t_W$ = 2692 s.\\}
\label{Fig 4}
\end{center}
\end{figure}

In a recent study, it was observed that the application of a strain of moderate magnitude enhanced the slow relaxation mode in a colloidal glass formed by polystyrene particles \cite{via_prl}. Based on the predictions of the trap model after a temperature step \cite{vin_glassy}, the authors argued that the observed behavior is suggestive of overaging, wherein a moderate shear can overpopulate the long time tail of the distribution of relaxation times. The soft glassy rheology (SGR) model \cite{sol_prl} predicts that under a moderate strain, the probability distribution function of the depths of the potential energy wells can split into two bumps corresponding to shallower and deeper well depths \cite{via_far}. Moderate strains can enhance the population of particles trapped in the deeper wells, making their escape more difficult and resulting in the slowing down of the system dynamics. More microscopically, the applied strain can alter the local curvature of the potential energy surface, which can lead to the disappearance of the intermediate barrier between two neighboring wells, such that the particle gets trapped in a deeper well \cite{mal_prl}. Since the energy minima visited under shear are different from those visited due to the aging process \cite{lac_prl}, the application of moderate strains can overage the system by relocating it to deeper energy minima. In previous experiments with aqueous foams, slow-down of the dynamics of the samples following the application of transient strains have been reported \cite{gopal_prl,cohen_prl}.

The slowing down of the slow relaxation process in our experiments for step strains between 1000\% and 12000\% is clearly reminiscent of earlier work on strain-induced dynamical slow-down. However, in addition to the enhanced sluggishness of the slow relaxation mode as represented by the fit to the stretched exponential function given by eqn. 1, we observe a decrease in the modulus of the system as the magnitude of the applied step strain is increased. While the former behavior demonstrates an overaging-like phenomenon, the latter behavior represents partial rejuvenation of the sample. We believe that this apparent contradiction is due to the way in which the distribution of particles in various energy wells determines the modulus of the system and the average relaxation time. If $b$ is the characteristic length-scale associated with the microstructure of the suspension and if $E_{i}$ is the depth of the energy well of the particle, the local modulus scale can be represented as: $G_{i}$ = $b^{3}E_{i}$ \cite{raljones_SM,awa_softm}. Correspondingly, the cage diffusion timescale associated with that particle is $\tau_{i} = \tau_{\circ}\exp(E_{i}/k_{B}T)$. For a nonergodic system, where $E_{i} >> k_{B}T$, the modulus of the material can be considered to be $G \sim \Sigma b^{3}E_{i}$. On the other hand, the relationship for the stretched exponential can be represented as: $\exp(-(t/\tau)^{B})$ = $\int_{0}^{\tau}\! \phi_{\tau,B}(\tau^{\prime})\exp(-t/\tau^{\prime})\, d\tau^{\prime}$ \cite{ruz_lang},  where $\phi_{\tau,B}$ is the appropriate distribution function [the integral can be considered as a summation over various $\tau_{i}$]. This suggests a possible scenario, wherein an applied step strain can overpopulate the long time tail, while rejuvenating the short time tail of the distribution of relaxation times. This would lead to a decrease in the modulus but an increase in the relaxation time as represented by the expression for the stretched exponential function. We believe that this is due to the the very complicated energy landscape of aqueous Laponite suspensions, wherein highly anisotropic particles of aspect ratio 25 with uneven charge distributions (negatively charged surfaces and weakly positive edges) are dispersed in a highly polar dielectric medium (water). Furthermore, an increase in the magnitude of the applied step strain may result in the alignment of these disc-like particles into a nematic order, as observed by Lele {\it et al.} \cite{lele_jor} in {\it in-situ} rheo-x-ray experiments. In our experiments, very high values of step strains are achieved over very short periods ($<$ 0.1 s). There is a strong possibility that strains of such high magnitudes, applied over these short time spans, induce dramatic changes in the complex microstructure of the system. However, without any direct evidence, it is difficult to speculate further on the specific microstructural details of Laponite suspensions on the application of step strains.

\begin{figure}
\begin{center}
\includegraphics[width=3.5in]{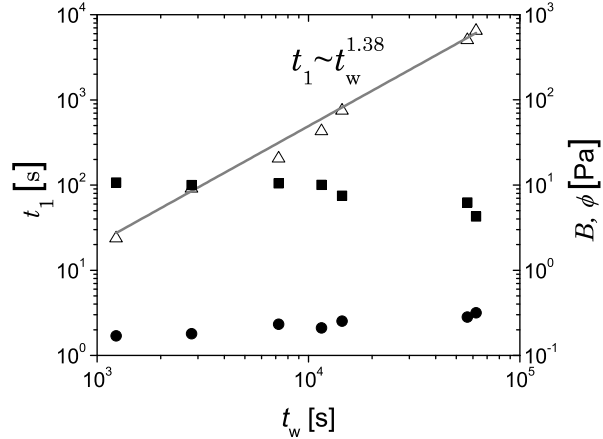}
\caption{The slow relaxation mode $t_1$ (open triangles), the intercept $\phi$ defined in eqn. 1 (filled squares) and the exponent B (filled circles) are plotted for different sample ages
following the application of a step strain of 300\%. \\}
\label{Fig 5}
\end{center}
\end{figure}

In order to observe the effects of sample age on the slow and fast relaxation
processes of Laponite suspensions, we next apply a constant step strain of amplitude
300\% to samples whose ages vary between 400 s and 30000 s. For
samples of age $t_W <$ 1000 s, the slower component of the stress relaxation data does not fit the KWW function, as the time over which the stress relaxes is much larger than the age of
the system. For $t_W >$ 1000 s, the data can be modeled by eq. 1 and the fitted
values of $t_{1}$, $\phi$ and B are plotted in fig. 5. As expected, we observe that the baseline of the fast process $\phi$ and the stretching exponent B are fairly insensitive to changes in
the sample age, while the slow timescale $t_1$ shows a power-law increase with
the age of the suspension. Our observations beautifully complement previous DLS and XPCS (x-ray photon correlation spectroscopy) studies that report
power-law aging of the slow relaxation process of Laponite suspensions \cite{ran_prl,bel_pre}. Interestingly, this power-law aging behavior is retained in our experiments even though they involve the application of moderately large step strain deformations. 

\section{Conclusions}

In conclusion, we have systematically studied the stress relaxation in aging Laponite suspensions on the application of step
strains. The stress, which relaxes in two steps, is
modeled by the KWW function, from which the slow
relaxation time scales are estimated. We observe that the slow mode, which is
independent of the magnitude of the step strain up to a strain of 1000\%, slows down as the shear strain is increased further. We believe that strains of high magnitudes applied over very short time
spans alter the complex microstructure of the Laponite suspensions considerably,
with the system exploring the deeper sections of the intricate energy landscape.  Furthermore, the slow relaxation time in our experiments shows a super-linear increase with sample
age. Our observations substantiate the complementary natures of stress relaxation and intensity autocorrelation functions in
describing the relaxational dynamics of aging systems. In contrast to optical techniques such as DLS that
measure microscopic particle dynamics, rheological measurements characterize bulk
quantities. Unlike optical techniques, nonlinear rheological techniques interfere
with the aging process and can provide additional insight into the effects of
deformation on the relaxation of aging systems.

\section{Acknowledgements}

The authors thank K. Sreejith for his assistance during the preliminary stages of
the experiments and the chemistry laboratory at the Raman Research Institute for
their help with sample storage. YMJ gratefully acknowledges financial support from
the Department of Atomic Energy, Government of India, under the BRNS young scientist
award scheme.

\end{document}